# Laser intensity noise suppression for space-borne gravitational wave mission

*Fan Li,[1] Xin Shang,[1] Zhenglei Ma,[1] Jiawei Wang,[1] Long Tian,[1,3,5] Shaoping Shi,[1]*
*Wangbao Yin,[2,3] Yuhang Li,[4] Yajun Wang,[1,3] Yaohui Zheng[1,3,6]*
*[1]State Key Laboratory of Quantum Optics and Quantum Optics Devices, Institute of Opto-Electronics, Shanxi University, Taiyuan 030006, China*
*[2]State Key Laboratory of Quantum Optics and Quantum Optics Devices, Institute of Laser Spectroscopy, Shanxi University, Taiyuan 030006, China*
*[3]Collaborative Innovation Center of Extreme Optics, Shanxi University, Taiyuan, Shanxi 030006, China*
*[4]State Key Laboratory of Precision Measurement Technology and Instruments, Department of Precision Instrument, Tsinghua University, Beijing 100084, China*
*[5]tianlong@sxu.edu.cn*
*[6]yhzheng@sxu.edu.cn*

**Abstract:**

Laser intensity noise is a main limitation of measurement and sensing mission represented by gravitational wave detection. We develop a noise decomposition model and design the core elements of the feedback loop based on the analysis results. We construct a fiber amplifier system with ultra-low intensity noise in the 0.1 mHz-1 Hz frequency band by the employment of an optoelectronic feedback loop that is specially designed, achieving a relative intensity noise below $6\times10^{-5}/\sqrt{\text{Hz}}$ @0.1 mHz-1 Hz. By precise control of the temperature, we improve the temperature and mechanical stability of fiber amplifier chain. Although the differential phase noise violates the LISA requirement in the 1 mHz-50 mHz frequency band, it meets the requirement at all other frequencies within the measurement range. The study provides experimental basis and technologies for precise measurement and sensing system at ultra-low frequency.

.

## 1. Introduction

The first direct observation of a gravitational wave in 2015 ushered a new era of the universe [1]. Further studies focus on a new generation of ground-based gravitational wave observatories (GWOs) with significantly increased sensitivity [2,3], and space-borne GWOs with extended frequency range [4-6], expecting to unlock the detection of more distant and fainter astronomical events. As a critical element of laser interferometer, laser intensity noise, and output power are two of the main limitations of the ground-based and space-borne GWOs. The requirements of gravitational wave (GW) detection continue to drive the laser performance boost. In addition, the development of low-noise, single-frequency laser will also propel the performance of coherent communication and remote detection forward [7-9].

Depending on the specific measurement and sensing mission, there are detailed requirements for laser specifications. Driven by ground-based GWOs, the performance of single-frequency lasers continues to improve, covering the output power, intensity noise and linewidth [10]. By employment of the master oscillator power amplifier (MOPA) configuration, the output power of the fiber amplifier has reached 1010 W at 1030 nm based on mode field manipulation technology [11]. Another independent topic is intensity noise suppression by a choice of seed laser and low-noise feedback control [12]. NASA Goddard Space Flight Center (GSFC) compared the influence of different kinds of master oscillator lasers, including fiber distributed Bragg reflector laser, Non-Planar Ring Oscillator (NPRO) laser, external cavity diode laser, and fiber ring cavity laser on laser intensity noise [13]. Following the comparison, the same group drew a conclusion that the micro-NPRO laser was an excellent candidate as a seed laser for the application of space-borne GWOs [14]. In addition, integrating with feedback control technology, the intensity noise of fiber amplifier was effectively suppressed at the low-frequency band [15-17]. In consideration of the influence of signal to noise ratio of feedback loop on ultimate intensity noise, innovative error signal extraction schemes, including high-power photodiode array [18], Optical AC coupling [19], and squeezing-assisted detection [20], have been demonstrated. Currently, the relative intensity noise (RIN) approximately approaches to $2\times10^{-9}/\sqrt{\text{Hz}}$ at Fourier frequency of 1 Hz to 10 kHz.

For the space-borne GWO mission [21], the ultra-low frequency band (0.1 mHz-1 Hz) corresponds to the characteristic signal frequency range of the primary scientific targets or gravitational wave sources in the LISA mission (e.g., supermassive black hole mergers, extreme mass ratio inspirals, etc.). In this band, relative intensity noise manifests as slow fluctuations in laser power, which can induce radiation pressure on the spacecraft's test masses [12] and consequently degrade detection sensitivity. Therefore, it is crucial to suppress the laser intensity noise within the 0.1 mHz-1 Hz range. The laser intensity noise suppression at the lower frequency band (0.1 mHz-1 Hz) is urgent to address. Moreover, the differential phase noise of fiber amplifier at 2.4 GHz should be evaluated to satisfy the clock synchronization requirement of the inter-satellite link. In fact, starting from the laser interferometer space antenna (LISA) mission [4], relevant works about laser development are still on the way to higher performance. Until mid-2022, a laser system with MOPA architecture was reported by GSFC, including two low-noise micro-NPRO seed lasers and a one-stage fiber amplifier [22]. The RIN of laser system was below $2\times10^{-4}/\sqrt{\text{Hz}}$ @0.1 mHz-1 Hz. Subsequently, the RIN of the laser system was further suppressed to below $4\times10^{-5}/\sqrt{\text{Hz}}$ at a lower frequency of 0.1 mHz [23], but at 0.5 mHz the RIN was $1.5\times10^{-4}/\sqrt{\text{Hz}}$, indicating that over the entire frequency band of 0.1 mHz to 1 Hz the RIN was not entirely below $1.0\times10^{-4}/\sqrt{\text{Hz}}$. Notably, the further improvement of the RIN faces a great challenge, especially at the ultra-low frequency 0.1 mHz.

In this paper, we demonstrate a fiber amplifier system with the RIN below $6\times10^{-5}/\sqrt{\text{Hz}}$ in the frequency range of 0.1 mHz to 1 Hz by employing an optoelectronic feedback loop, which meets the requirement for space-borne GWOs. According to feedback control theory, we construct a noise decomposition model, quantitatively obtaining the specification requirement for core elements (voltage reference, photodetector etc.) in the loop. Further, we design independently these core elements, integrate these electric units with fiber amplifier, and construct an ultra-low intensity noise fiber amplifier system. Exploiting a passive-stabilized scheme, the differential phase noise is suppressed to below $5\times10^{-2}\text{rad}/\sqrt{\text{Hz}}$ at 0.1 mHz. Although it does not meet the LISA requirement in the 1 mHz-50 mHz band, it remains below $3\times10^{-4}\text{rad}/\sqrt{\text{Hz}}$ in the 100 mHz-1Hz range. These design schemes, results, technical intricacies and the low-noise single frequency fiber amplifier system provide experimental basis and technologies for precise measurement system at ultra-low frequency.

## 2. Power stabilization noise model

MOPA structure is selected for our low-noise fiber laser amplifier system, which adopts large mode field area double cladding ytterbium doped gain fiber as well as cladding pumping technology. Following the previous studies [24-26], the RIN of laser amplifier at a low frequency band mainly comes from its pump noise, not the noise of seed laser. Therefore, we employ the pump driver of laser amplifier as the actuator of the feedback loop.

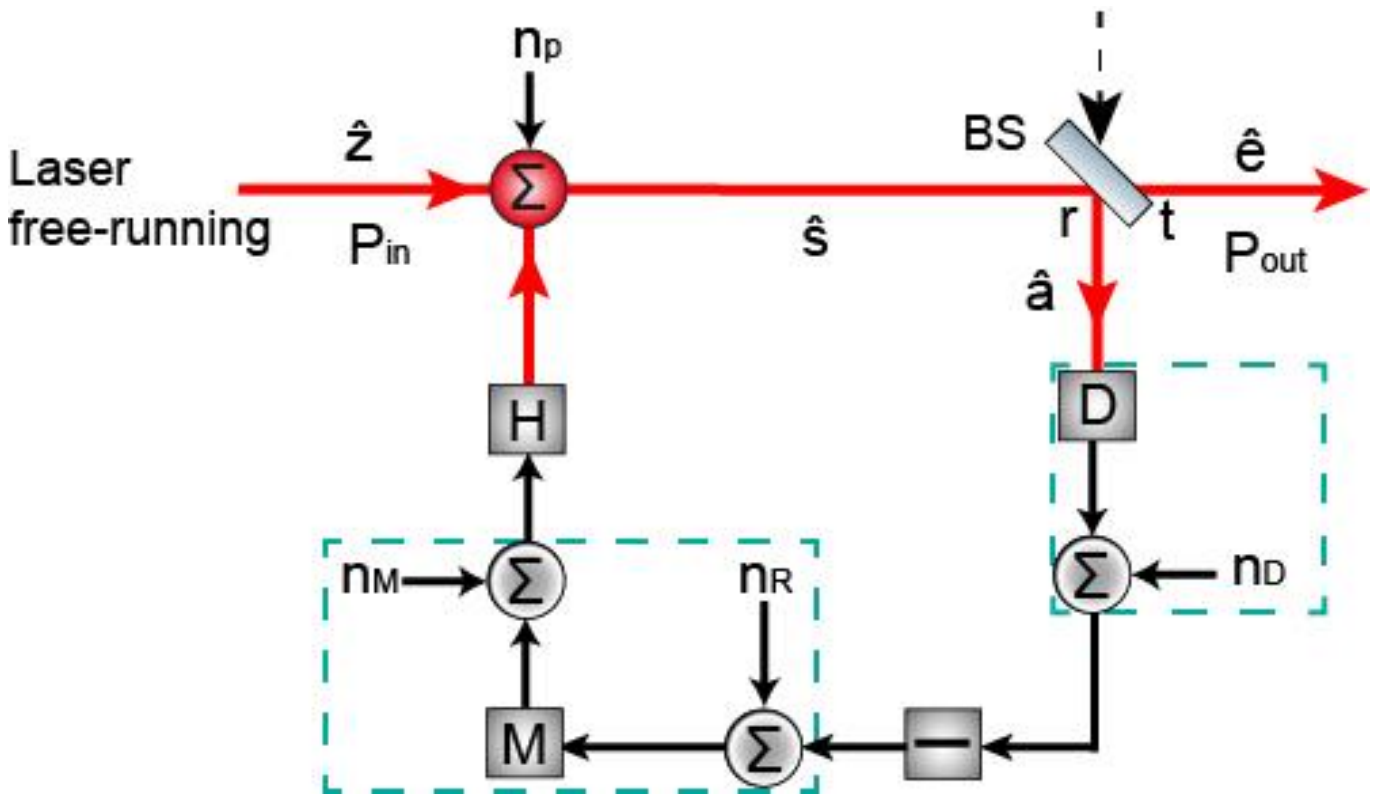

Fig.1 Schematic representation of power stabilization noise model.

The noise suppression scheme is based on the active feedback loop, as shown in Fig 1. $\hat{z}$ is the input laser field and the input power is denoted by $P_{\text{in}}$. The technical noise introduced by the fiber amplifier is included in $\hat{z}$. The pump noise introduced by the actuator in the active control loop is denoted by $n_P$. After the actuator, a beam splitter (BS) extracts a small fraction of the light from the field $\hat{s}$ and directs it into the photodetector. Here, the BS is a lossless beam splitter, i.e. $r^2 + \text{t}^2 = 1$ ($r^2$ and $\text{t}^2$ represent the power reflectivity and power transmissivity, respectively). The signal measured by the photodetector is applied to the actuator through feedback to cancel the power fluctuation signal. $\hat{a}$ is the in-loop laser field, $\hat{e}$ is the out of loop laser field and the output power is denoted by $P_{out}$. In addition, the feedback loop gain is represented by the detector gain D, the controller gain M, and the actuator gain H, respectively, i.e., G = HMD. $n_D$ and $n_R$ represent the fluctuations from the detector and reference in the feedback loop, respectively. $n_M$ is the fluctuation introduced by the controller. The amplitude quadrature of the output noise field is expressed in Eqs (1)-(2):

$$\hat{s} = \hat{z} + n_P + Hn_M + HMn_R - HM\text{n}_D - HMD\hat{a} \tag{1}$$

$$\hat{e} = \frac{t\left(\hat{z} + n_P\right)}{1 + rG} + \frac{tH}{1 + rG} n_M + \frac{tG}{1 + rG}\frac{n_R}{D} - \frac{tG}{1 + rG}\frac{n_D}{D} \tag{2}$$

Assuming that all inputs to the system are uncorrelated, the power spectral density quadrature to the amplitude of the output field can be calculated in Eq (3). where $\gamma_{n_R}^2$, $\gamma_{n_M}^2$, and $\gamma_{n_D}^2$ represent the electronic noise contribution in $\text{V}^2$/Hz, respectively. $PSD_{tech}$ is the sum of actuator noise and input field noise, and so:

$$PSD_{\hat{e}} = \frac{t^2 PSD_{tech}}{\left|1 + rG\right|^2} + \frac{t^2 \left|H\right|^2 \gamma_{n_M}^2}{\left|1 + rG\right|^2} + \frac{t^2 \left|G\right|^2}{\left|1 + rG\right|^2}\frac{\gamma_{n_R}^2}{\left|D\right|^2} + \frac{t^2 \left|G\right|^2}{\left|1 + rG\right|^2}\frac{\gamma_{n_D}^2}{\left|D\right|^2} \tag{3}$$

Next, defining the RIN of the output field $\hat{e}$ [27], where the output field's carrier power is given by $P_{out} = t^2 P_{in}$. The RIN of the output field is expressed in Eqs (4)-(5) :

$$RIN^2\hat{e} = \frac{2\hbar\omega}{P_{out}} PSD_{\hat{e}} = \frac{2\hbar\omega}{t^2 P_{in}} PSD_{\hat{e}} \tag{4}$$

$$RIN^2\hat{e} = \frac{RIN^2{}_{tech}}{|1+rG|^2} + \frac{2\hbar\omega}{P_{in}} \frac{|H|^2 \gamma_{nM}^2}{|1+rG|^2} + \frac{2\hbar\omega}{P_{in}} \frac{|G|^2}{|1+rG|^2} \frac{\gamma_{nR}^2}{|D|^2} + \frac{2\hbar\omega}{P_{in}} \frac{|G|^2}{|1+rG|^2} \frac{\gamma_{nD}^2}{|D|^2} \tag{5}$$

Where $\hbar$ is the Planck constant, $\omega$ is the optical angular frequency. From Eq (5), the RIN comes from the contribution of residual technology noise, controller noise, voltage reference noise and photodetector electronics noise, respectively. For large loop gains (G>>1), the influence of the residual technical noise $RIN^2{}_{tech}$ and the controller noise $\gamma_{nM}^2$ on the RIN can be neglected, the voltage reference noise $\gamma_{nR}^2$ and photodetector electronics noise $\gamma_{nD}^2$ is main limitation of the RIN. The conclusion points out our direction of the experimental realization.

## 3. Experimental setup

Fig. 2 shows the experimental setup of our laser amplifier system. The laser amplifier consists of laser amplifier unit (LAU), active stabilization unit (ASU), and electrical acquisition and control unit (EACU). The LAU and ASU are independently placed on an insulation house with active temperature control.

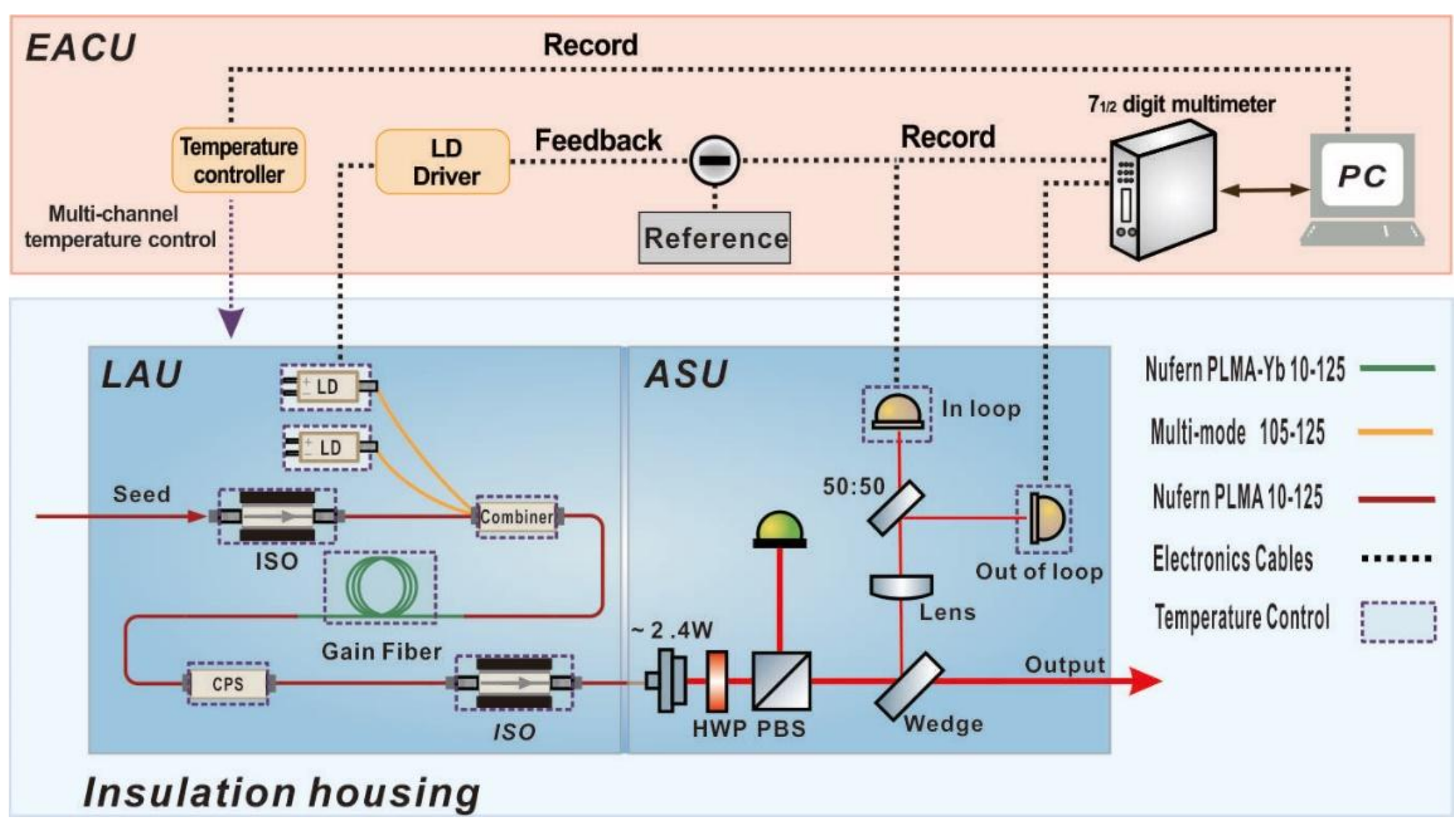

Fig.2 Experimental setup of low-noise single-frequency laser amplifier system. ISO: Isolator: LAU: laser amplifier unit; ASU: active stabilization unit; EACU: electrical acquisition and control unit.

A NPRO with a center wavelength of 1064 nm and an output power of 500 mW serves as its seed source (not shown in the Fig. 2), and after free space attenuation, 50 mW of optical power is injected into the laser amplifier. Two series-connected 976 nm laser diodes (Aerodiode, 976 LD, 10 W-105 μm) are employed as the pump source for the laser amplifier, which is injected into the gain fiber with the same direction as the seed source after passing through a (2+1) × 1 polarization-maintaining combiner. The gain fiber in the amplifier is a 2.5 m ytterbium-doped double-clad fiber (Nufern, PLMA-YDF-10/125) with 10 μm core diameter, 125 μm pump cladding diameter, and numerical apertures of 0.075 and 0.46 for core and pump cladding, respectively. We use one-component silicone to secure the gain fiber to a copper heat sink with U-shaped grooves etched into it, and precisely control the temperature of the copper heat sink. To reduce the influence of the residual 976 nm pump light on the downstream main laser, we used a cladding pump stripper (CPS) at the end link of the gain fiber. Furthermore, a polarization-maintaining fiber optic isolator (ISO) is installed in the output of the seed laser and the output of the amplifier respectively, which significantly reduces the influence of backscattered light on the output signal light. Another advantage of the ISO is that it eliminates parasitic interference, thereby improving the low-frequency performance of the laser amplifier. These elements mentioned above are installed inside the LAU.

The output of our fiber amplifier is collimated by a fiber collimator to produce a 2 mm beam in diameter. The laser first passes through a half-wave plate with a polarizing beam splitter (PBS) to obtain a horizontally polarized laser. A wedge splitter is placed after the PBS, and an integrating sphere power meter is used to detect the laser at the transmission end. The laser at the reflection end is focused by a lens with a focal length of 100 mm and then split into two beams by a non-polarizing 50:50 beam splitter. Two custom-made low-noise photodetectors are employed as the in-loop and out-of-loop signal acquisition, respectively. The substrate of the 50:50 beam splitter is fused silica, which has a thermal expansion coefficient of $0.55\times10^{-6}$/K. To minimize ghosting, the 50:50 beam splitter has a 30 arcmin wedged back surface. These mentioned free-space devices, including fiber collimator, PBS, beam splitters and low-noise photodetectors, are positioned inside the ASU. The electrical component is shown in the upper part of Fig. 2, named as EACU. The EACU includes the voltage reference, current driver, temperature controller etc., which is detailed in Section 5.

## 4. Characterization of free-running laser amplifier

In free-running mode, we measure the output power of the laser amplifier as a function of the pump power, as shown in Fig. 3(a). The output power increases linearly with the pump power with a slope efficiency measured of 56.45 %. In consideration of the damage threshold of the ISO, the laser amplifier operates at 2.4 W for extended periods. There is no obvious stimulated Brillouin scattering phenomenon at the operating power, and the polarization extinction ratio is above 28 dB. The laser spectrums of the seed laser and laser amplifier are given in Fig. 3(b). The resolution and scanning range of the optical spectrum analyzer (Yokogawa AQ6370D) are set to 0.02 nm and 150 nm, respectively. As observed, the center wavelength of the seed laser is 1064.45 nm with an optical signal-to-noise ratio (OSNR) of ~72 dB and no amplified spontaneous emission (ASE) is generated. Moreover, the additional linewidth induced by the fiber amplifier is negligible [16], and its linewidth remains ~ 2 kHz.

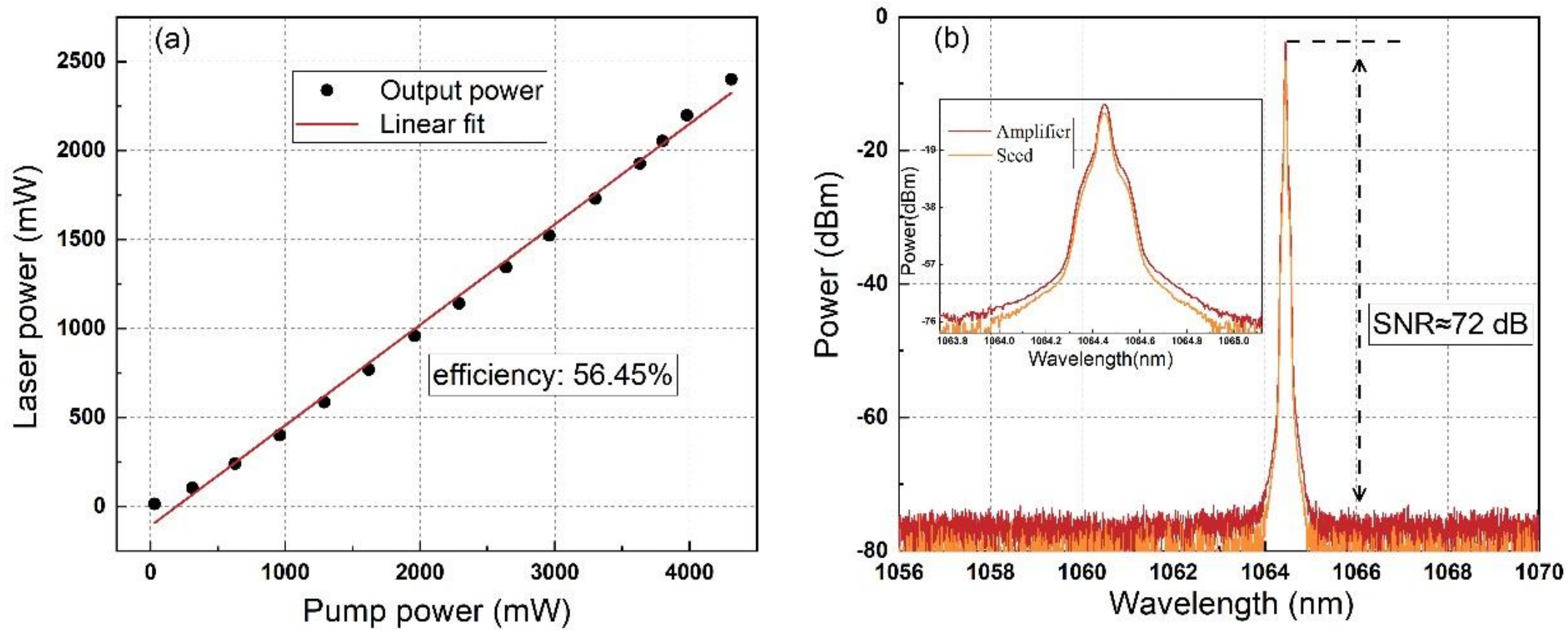

Fig.3 (a) Output power of the laser amplifier versus pump power. (b) Laser spectra for seed laser and power amplifier.

Here, we mainly focus on the RIN suppression of the laser amplifier. Although the frequency noise is primarily dependent on the seed laser, we also measure the frequency noise of the laser amplifier in order to make a global investigation into the application of space-borne GWOs. Utilizing the optical heterodyne beat-frequency technique [28], we measured the frequency noise of the free-running laser amplifier. Two NPRO lasers with identical cavity lengths, manufacturing processes, and packaging temperature controls were employed to form the heterodyne beat frequency system. By tuning the lasers via temperature control and PZT adjustments, the heterodyne beat signal was brought within the bandwidth of the photodetector (Thorlabs DET08CFC). A frequency counter (Keysight 53220A) was then used to continuously record the beat signal at a sampling rate of 30 Hz over a 3-hour duration, enabling frequency noise measurements as low as 0.1 mHz. Subsequently, the acquired data were processed using an Fast Fourier Transform algorithm, which enabled the calibration of the reference laser's frequency noise. Finally, heterodyne beat frequency measurements were conducted between the reference laser and the laser under test after amplification by the fiber amplifier. Based on the noise superposition principle (summation of squares), the frequency noise spectrum of the free running laser amplifier under test was obtained, as shown in Figure 4(a). The result meets the scientific requirement that the LISA mission claims for the frequency noise of the free-running laser [21,29].

The space-borne GWO needs to transfer clock information among its three spacecraft in the form of GHz phase-modulation sidebands [30,31]. For this reason, the phase noise introduced by the optical chain between the carrier and sideband must be as low as possible [32,33]. The differential phase noise of laser amplifiers is commonly used as a performance indicator. According to the existing analysis, the differential phase noise originates from the thermal load on the active fiber, thermo-mechanical induced birefringence, nonlinearities in fibers and temperature fluctuations [34,35]. Especially at ultra-low frequency, the temperature and mechanical fluctuations are the main limitation of the differential phase noise. Here, we construct a high-stability, stress-free stove that has a temperature fluctuation of less than 5 mK peak-to--peak, for the fiber amplifier chain to isolate environmental disturbance. Via the passive-stabilized scheme, the differential phase noise is effectively suppressed. The measurement setup for the differential phase noise has been constructed based on reference [36]. The outputs from two NPRO lasers are combined using a 3 dB coupler (2\text{x}2). Before and behind the amplifier, optical signals are detected by fiber-coupled photodetectors (Thorlabs DET08CFC). Before the fiber amplifier, the detected signal is split into two paths: one path is mixed with a local oscillator (RIGO DSG3030, frequency range: 9 kHz-3 GHz) and then fed into an offset phase locked loop (VESCENT D2-135) to lock the differential frequency of the two lasers at 2.4 GHz; the other path, together with the signal detected after the fiber amplifier, is mixed with a second local oscillator (Keysight M5181B, frequency range: 9 kHz-3 GHz) at a slightly offset frequency (2.4 GHz + 1.5 MHz) and subsequently low-pass filtered. The resulting two 1.5 MHz signals are then input into the Moku:lab (utilizing its phase meter function) to record phase difference fluctuations. To minimize temperature effects, all optical and electronic components are temperature controlled via a stabilized cold plate and an insulated enclosure. Additionally, the fiber amplifier employs the LAU unit from Fig.2 and does not implement active power stabilization. Fig. 4(b) shows the measurement result of the differential phase noise. The gray curve is the noise floor of the phasemeter (Moku: lab), which is obtained by recording the phase difference of the same sinusoid signal. By replacing the gain fiber with a short, passive fiber, we record the noise floor of the measurement setup, shown with the blue curve in Fig. 4 (b). The red curve of Fig. 4 (b) is the measurement result of the differential phase noise of the fiber amplifier at the output power of 2.4 W, not subtracting the dark noise. In the future, we expect to suppress the differential phase noise by optimizing the doping concentration and size of gain fiber, and the stove construction, to satisfy the LISA mission.

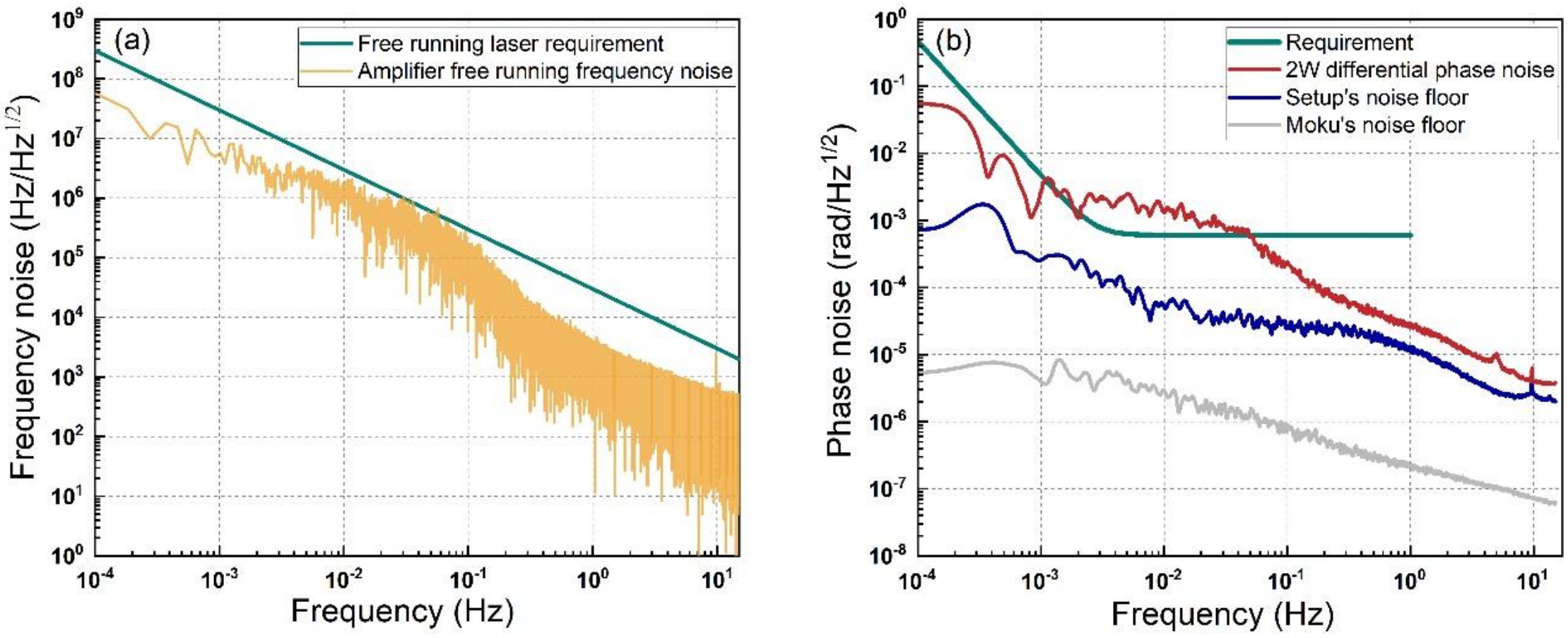

Fig.4 (a) Frequency noise and (b) differential phase noise of the fiber amplifier in the 0.1 mHz-15 Hz frequency band.

## 5. Characterization laser amplifier with power stabilization

According to the power stabilization noise model of Section 2, we develop an active feedback control system to suppress the RIN of laser amplifiers in ultra-low frequency bands. We choose the driver of pump laser as an actuator of the feedback loop, which is a general technique [37-39]. As is shown by the EACU in Fig. 2, a photodetector converts a fraction of laser power disturbances into an electrical voltage signal. The signal is subtracted from the reference voltage to obtain the feedback loop error signal, which is amplified and filtered by the servo controller to manipulate the power supply of the pump laser. In our system, the unity gain frequency of the control loop is about 5 kHz. For our experiment in the 0.1 mHz-1 Hz frequency band, we should pay more attention to the noise coupling of the components in the feedback system, the 1/f noise, and the temperature coefficient of the devices, rather than the feedback bandwidth, which distinguishes the high-frequency range power stabilization experiments to some extent.

Based on the analysis in Section 2, we first upgrade the photodetector. Thermoelectrically cooled 5 mm InGaAs photodiodes from Hamamatsu's G12180-150A are used, which enables both low-noise and low dark current. And the G12180-150A has an internally integrated Thermoelectric Cooler (TEC) that keeps the temperature and responsivity of sensitive area stable. It should be noted, however, that while temperature stabilization can help reduce differential noise between the two detectors to some extent, there will still be residual differences in responsivity between the two photodetectors. These residual differences can affect the measurement accuracy and overall system performance. The photodiode operates in photovoltaic mode, which provides lower dark noise without reverse bias [40]. As is shown by the orange curve in Fig. 5 (a), the measured photodetector with an electronic noise $\gamma_{nD}$ of less than $9.5\times10^{-7}$ V/√Hz @0.1 mHz-1 Hz. The noise spectrum is relatively flat throughout the frequency range and no significant 1/f noise is observed. This is largely due to the zero-drift operational amplifier (Analog Devices LTC1151) used in the TIA circuit, which has a unique chopping technique that effectively reduces the effects of 1/f noise. In addition, these excellent performances also benefit from a low noise power supply and a precision low temperature coefficient metal foil resistor.

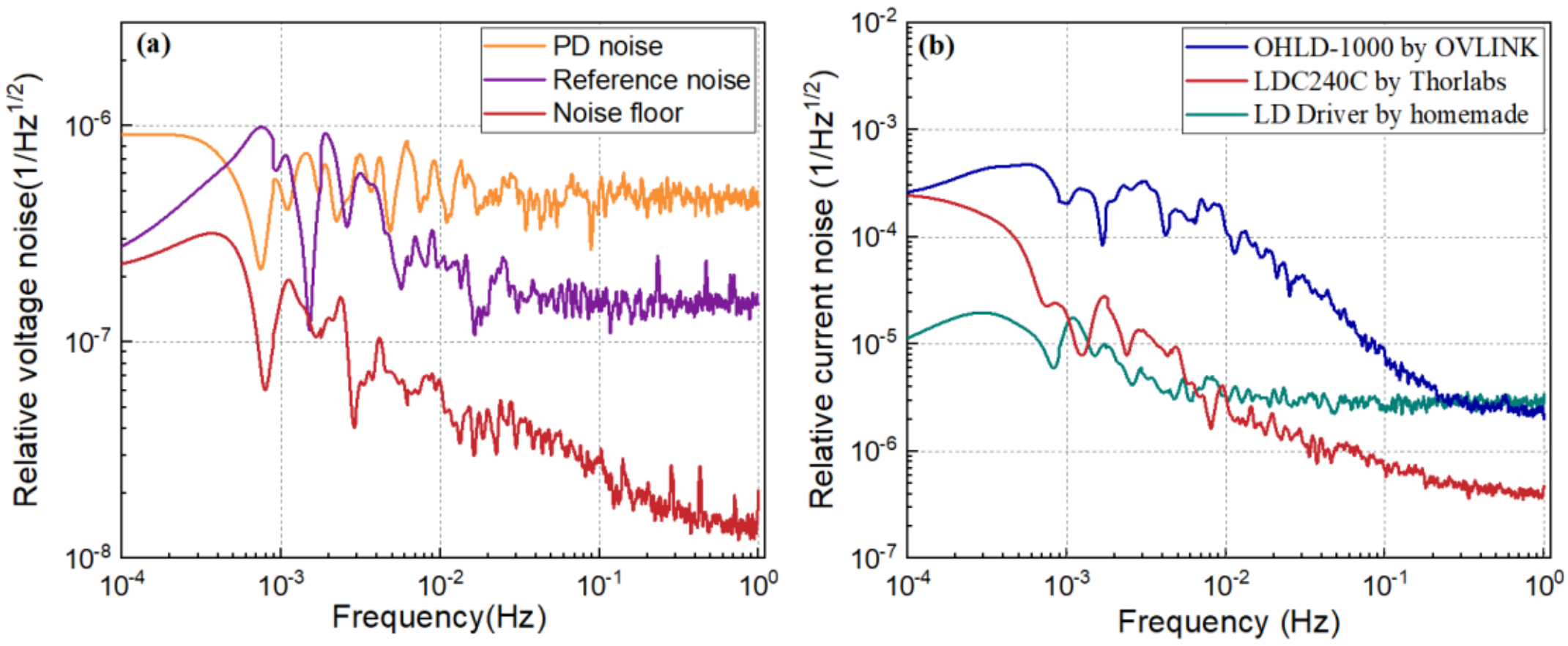

Fig.5 (a) Voltage reference performance and photodetector dark noise. (b) Comparison of current noise spectral density between homemade and commercial pump drivers.

The voltage reference is another limitation of the RIN of the laser amplifier. The reference voltage provided by the ultra-precision reference (Linear Technology LTZ1000) has an extremely low temperature coefficient, typically less than 0.05 ppm/°C (up to 0.01 ppm/°C when equipped with an external temperature controller), meaning that temperature variations have a minimal effect on the output voltage. The purple curve in Fig. 5 (a) shows the relative voltage noise spectral density $\gamma_{nR}$of the voltage reference, which is superior to 2×10$^{-6}$ V/√Hz in the range of 0.1 mHz-1 Hz. The indicator is two orders of magnitude better than of the RIN requirement, which meets the design requirement. In ideal condition, the influence of the residual technical noise $RIN^2{}_{tech}$ and the controller noise $\gamma_{nM}^2$ on the RIN can be neglected. However, the actual feedback loop does not have an infinite gain factor. Therefore, it was beneficial to upgrade the current driver. Integrating the pump driver with controller and actuator circuits on board, we greatly reduce ambient noise coupling, further reduce the residual technical noise $RIN^2{}_{tech}$ and the controller noise $\gamma_{nM}^2$ of the laser amplifier system in free-running mode. It is worth noting that our design for the laser driver focuses on the low-frequency performance. Fig. 5 (b) shows the comparison between the custom-made driver and commercial ones. The noise level of the custom-made driver in the range of 0.1 mHz-5 mHz is significantly lower than that of other commercial drivers, which is relatively flat throughout the concerned frequency range.

The red curve of Fig. 6 shows that the RIN of the free-running laser amplifier with below 4×10$^{-3}$/√Hz @0.1 mHz-1 Hz, which is driven by custom-made laser driver. In Eq.5, the detector gain $D$ can be expressed as$D = k\sqrt{2\hbar\omega r^2 P_{in}}$, where $k$ is the photodetector calibration factor ( $k$ =27300V/W), which is related to the photodiode responsivity, quantum efficiency and transimpedance resistance. Although it is difficult to accurately measure the loop gain at ultra-low frequencies, it can be roughly estimated based on the DC responses of each component. In our system, the photodetector's DC response is 27300 V/W; the controller's DC response is 1 A/V; the actuator's DC response is 0.9 W/A; and the pump-to-amplifier DC response is 0.564, which yields a loop gain of approximately 14000. Subsequently, we choose G = 14000 (assumed to be frequency independent) for calculation, the expected out-of-loop RIN is an uncorrelated superposition of the residual technology noise, the controller noise, the photodetector electronics noise and the reference noise. Instituting these measurement results of each independent element into Eq. 5, we obtain the expected out of loop RIN noise, shown with orange curve in Fig. 6. The blue curve shows the out-of-loop performance of the laser amplifier with RIN of below $6\times10^{-5}/\sqrt{Hz}$@0.1 mHz-1 Hz. According to the Eq. 5, the in-loop noise is an uncorrelated superposition of the residual technology noise and the photodetector noise, which is shown by the black curve in Fig.6. It is necessary to note that that out-of-loop noise almost matches the expected value in the 200 mHz-1 Hz range, but at low frequencies (< 200 mHz) is not as expected. We believe that the observed discrepancy is mainly due to the influence of 1/f noise at extremely low frequencies, which may originate from long-term temperature drifts affecting the stability of the entire feedback system. Since our theoretical model did not account for this additional 1/f noise, inconsistencies between the theoretical predictions and the experimental data appear below 200 mHz. Furthermore, in the 10 mHz-1 Hz band, the out-of-loop noise approximately overlaps with the in-loop one. At the lower frequency bands, the out-of- loop noise is inferior to the in-loop one. This phenomenon may be attributed to the residual temperature noise of the Photodiode and the "ghosting" effect resulting from the detector window, photodetector sensitivity fluctuation from beam pointing.

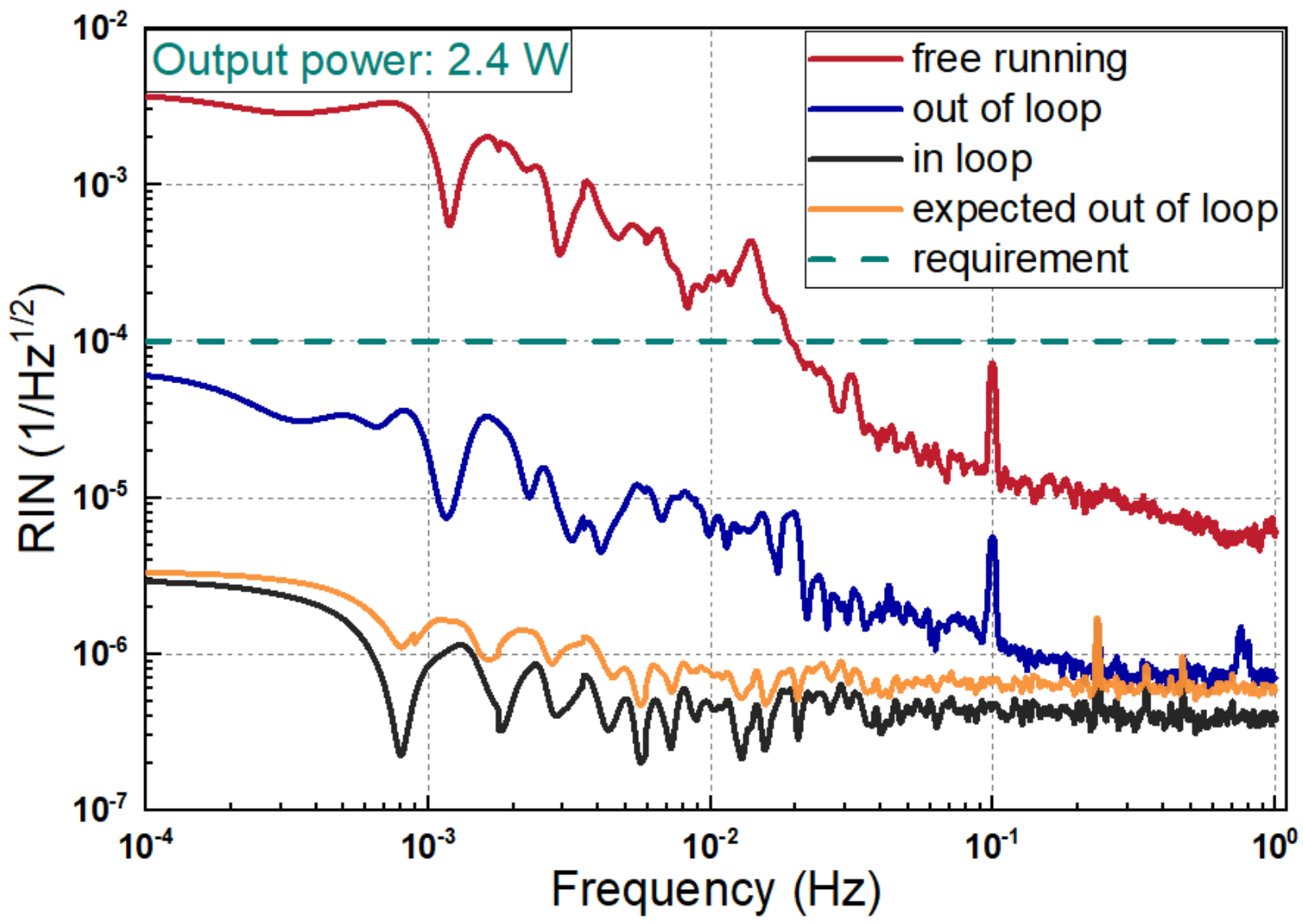

Fig.6 Relative intensity noise of the laser amplifier.

## 6. Conclusion

We have demonstrated a fiber amplifier system with 2.4 W output by the employment of an optoelectronic feedback loop and precise temperature control. Starting from feedback control theory, we quantify the specification requirement for core elements. By designing the low-noise voltage reference, photodetector, and laser driver in the 0.1 mHz–1 Hz frequency band, we have experimentally achieved a RIN below $6\times10^{-5}/\sqrt{Hz}$@0.1 mHz-1 Hz, which meets the requirement for space-borne GWOs. The experimental results deviate from the theoretical expectations below 200 mHz, primarily due to the model's omission of the impact of 1/f noise at ultra-low frequencies. By precisely controlling the temperature, the differential phase noise is suppressed to below $5\times10^{-2}$ rad/$\sqrt{Hz}$ at 0.1 mHz. Although it does not meet the LISA requirement in the 1 mHz-50 mHz band, it remains below $3\times10^{-4}$rad/$\sqrt{Hz}$ in the 100 mHz-1 Hz range. The work points out our direction of further performance improvement and offers essential technologies for the noise suppression of laser amplifier at low-frequency band.